\documentclass[final,english]{bullsrsl}

\usepackage[latin1]{inputenc}

\usepackage[T1]{fontenc}

\usepackage{natbib} 

\usepackage{graphicx}
\usepackage[normalem]{ulem}

\begin{document}
\title{Exploring pulsational instabilities in the O-type supergiant HD\,152249}



  \author[affil={1}, corresponding]{Subharthi}{Dasgupta}
  \author[affil={1}]{Abhay Pratap}{Yadav}
  \author[affil={1}]{Sugyan}{Parida}
  \author[affil={2}]{Santosh}{Joshi}

\affiliation[1]{\small Department of Physics and Astronomy, National Institute of Technology, Rourkela-769008, India}
\affiliation[2]{\small Aryabhatta Research Institute of Observational Sciences, Manora Peak, Nainital - 263002, India}


\correspondance{dasguptasubharthi@gmail.com}


\maketitle

\begin{abstract}
Variability in O-type supergiants is a well known phenomena although their origin is not fully understood. In very massive, luminous stars, strange mode pulsations have been suggested to play a role in the variability. Here, we present a preliminary theoretical study of O-type supergiant models corresponding to the observed parameters of the star HD 152249, which shows line profile variability. Non-adiabatic linear stability analysis with respect to radial perturbations is carried out for the considered models of this star. Instabilities with growth rates of the order of dynamical timescale are found, with characteristics suggestive of strange modes. Non-linear simulations for selected models indicate that the instabilities could lead to significant envelope inflation, finite amplitude pulsation and may contribute to enhanced mass loss.
\end{abstract}

\keywords{Massive stars, Supergiants, O-type stars, HD 152249, Instabilities in stars}




\section{Introduction} \label{intro}

Massive O-type stars, including supergiants, are known to exhibit significant photometric and spectroscopic variabilities \citep{fullerton1996,kambe1997,markova2005,prinja2006,blomme2011,howarthstevens2014,rauw2015,buysschaert2017}. The variability has been attributed to radial and nonradial stellar pulsations for some of these stars. These include $\zeta$ Oph \citep{walker2005}, HD 93521 \citep{howarthreid1993,rauw2015}, $\lambda$ Cep, $\xi$ Per \citep{dejong1999} and HD 46202 \citep{briquet2011}. However, the class of O-type supergiants has not been sufficiently explored. Theoretical studies show that models of massive, luminous stars are susceptible to strange mode instabilities \citep{kiriakidis1993, saio2011} that may lead to pulsation and associated mass loss \citep{yadav2017}. \cite{fullerton1996} found that the confirmed and suspected pulsating O-type stars in their sample lies within the domain of instability found by \citet{kiriakidis1993}. \citet{godart2011} found strange modes to be excited in stellar models in a region of the Hertzsprung-Russell (HR) diagram containing O-type stars observed in the CoRoT mission. Motivated by such studies, we present a preliminary theoretical investigation of instabilities and their consequences in models of O-type supergiants. For the purpose of our study, we consider the O-type supergiant HD 152249 (OC9Iab) which has been known to exhibit significant line profile variations \citep{gosset2009}. 

A linear non-adiabatic stability analysis with respect to radial perturbations is followed by non-linear simulations for selected unstable models. The stellar models and the methods for the stability analysis and non-linear simulations are briefly described in Section \ref{models_methods}. Results of the linear analysis and non-linear simulations are given in Sections \ref{lna} and \ref{nonlinear} respectively. The discussion and conclusions follow in Section \ref{conclusion}. 

\section{Models and Methods} \label{models_methods}

In order to construct the models resembling HD\,152249, we adopt the effective temperature T$_\mathrm{eff}$ = 31500 K and luminosity of $\log \left(\frac{L}{L_\odot}\right)$ = 5.59 from \citet{markova2018}. To account for the uncertainty in mass, a sequence of envelope models are constructed with the temperature and luminosity fixed and the mass ranging from 15 to 40 M$_\odot$, which includes the value of the star's spectroscopic mass (25.7 M$_\odot$). Assuming a solar chemical composition (X = 0.70, Y = 0.28, Z = 0.02), the stellar structure equations are integrated from the photosphere to an interior point having temperature of $10^7$ K where the nuclear burning does not occur. 
The Stefan-Boltzmann law and photospheric pressure are taken as initial conditions for the integration. Previous investigations \citep[e.g.,][]{glatzel1993} have found that the instabilities in massive stars are generally present in their envelopes while the stellar core is not significantly affected by the pulsation. Accordingly, the adopted cut-off temperature excludes the core, considering only the stellar envelope. 
Rotation and magnetic fields are not taken into account. The onset of convection is defined by Schwarzschild's criterion. Standard mixing length theory \citep{bohmvitense1958} with a mixing length parameter of 1.5 is used to describe the convection. Opacities are provided by the OPAL opacity tables \citep{rogersiglesias1992,iglesias1996, rogersswenson1996}. 

The models are subjected to a non-adiabatic linear stability analysis with respect to radial perturbations according to the prescription of \citet{gautschy1990b}. Convection is treated in the frozen-in approximation \citep{baker1965}, where the perturbation of the convective flux is neglected. This is expected to hold if the major fraction of the energy in the stellar envelope is transported by radiation. In fact, for the considered models, not more than 30\% of the total luminosity is transported by convection. The solution of the resultant boundary eigenvalue problem provides a set of complex eigenfrequencies for each model. The real part of the eigenfrequency ($\sigma_r$) is associated with the pulsation frequency while the imaginary part indicates excitation ($\sigma_i<0$) or damping ($\sigma_i>0$). The eigenfrequencies are normalized by the global free fall time ($\sqrt{\frac{R^3}{3GM}}$, $R$
being the stellar radius, G the gravitational constant and $M$ the stellar mass) of the associated model. 

In order to understand the final outcome of the instabilities they must be followed into the non-linear regime. This is done on the basis of a numerical scheme as described in \citet{grott2005}. As in the linear analysis, convection is treated in the frozen-in approximation.
The code is provided an initial model in hydrostatic equilibrium. It picks up the physical instability from numerical noise and enters a linear phase of exponential growth following which it saturates in the non-linear regime. Since the kinetic and acoustic energies, which are of interest in pulsation, are several orders of magnitude smaller than the gravitational and internal energies, they must be determined with high accuracy. To do so, the energy balance must be satisfied intrinsically, which is ensured by using a fully conservative scheme.

\section{Results}

\subsection{Linear Stability Analysis} \label{lna}

\begin{figure}
\centering

\begin{minipage}{0.48\textwidth}
    \centering
    \includegraphics[width=\linewidth]{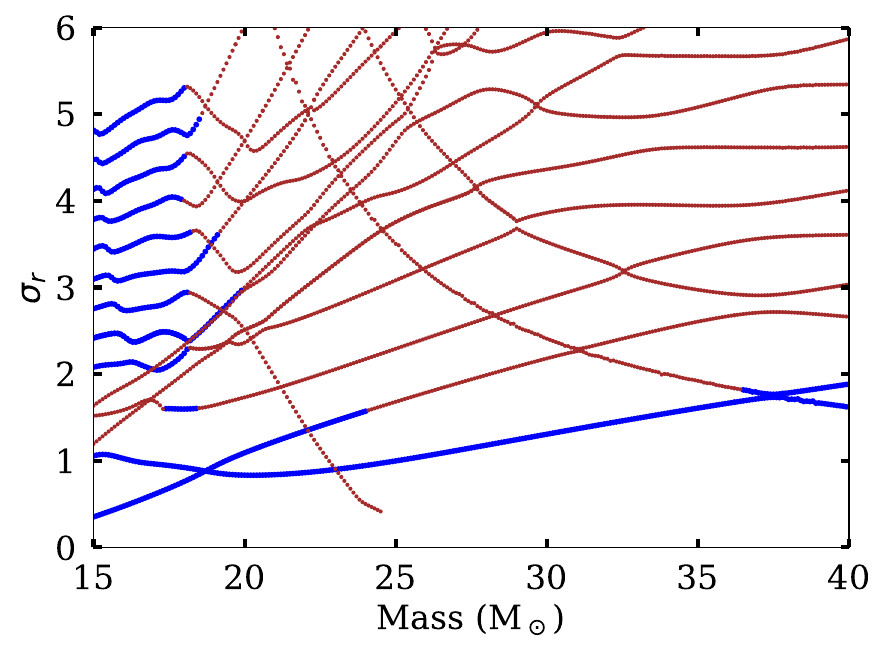}
\end{minipage}
\hfill
\begin{minipage}{0.48\textwidth}
    \centering
    \includegraphics[width=\linewidth]{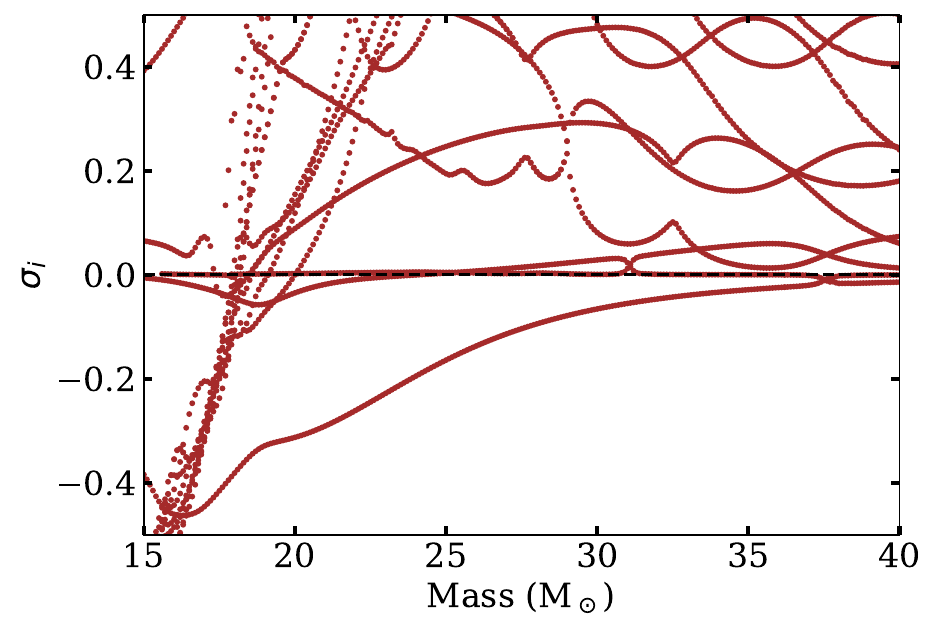}
\end{minipage}
\caption{Results of the linear stability analysis for models of HD 152249. The real (left) and imaginary (right) parts of the eigenfrequencies are given as a function of mass. The negative imaginary part indicates unstable modes, which are correspondingly shown by thick blue lines in the real part.}
\label{fig1:lna}
\end{figure}

The results of the linear stability analysis are presented in Fig.\ref{fig1:lna} in the form of a modal diagram \citep{saio1998} where the real and imaginary parts of the dimensionless eigenfrequencies are plotted as a function of the stellar mass. The thick blue lines denote the excited modes i.e. modes with a negative imaginary part. All the models are found to be unstable in at least one low order mode.  Another mode, which is unstable only below 25 M$_\odot$, crosses this mode close to 18 M$_\odot$. For models below 20 M$_\odot$, several modes are found to be excited. In between 25 and 35 M$_\odot$ model, only a single mode is excited. For masses above approximately 35 M$_\odot$, an additional mode is found to be unstable. Below 35 M$_\odot$, the corresponding mode is damped and appears to increase significantly as the mass decreases. In general, the number and growth rate of the instabilities are found to increase with decreasing mass. Stability analysis of the considered models with a deeper inner boundary (5$\times$10$^7$ K) showed that the eigenfrequencies and growth rates do not change significantly. Thus, the modes are insensitive to the exact position of the inner boundary provided it is located sufficiently deep inside the star.
The periods associated with the unstable modes are found to range from 0.35 to 5 d.

\begin{figure}
\centering

\begin{minipage}{0.48\textwidth}
    \centering
    \includegraphics[width=\linewidth]{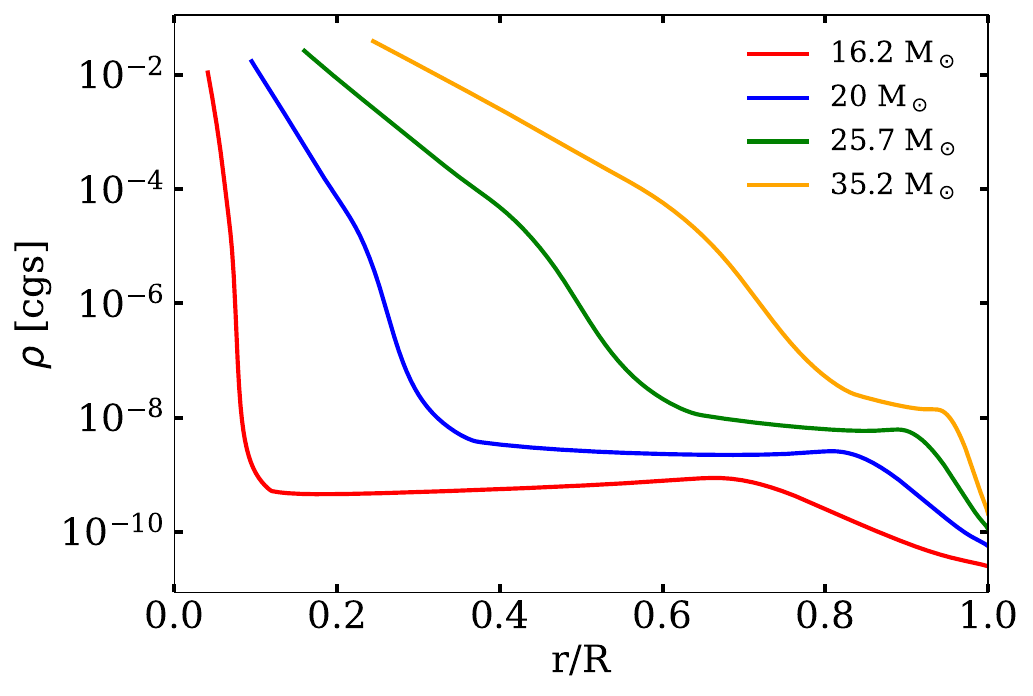}
\end{minipage}
\hfill
\begin{minipage}{0.48\textwidth}
    \centering
    \includegraphics[width=\linewidth]{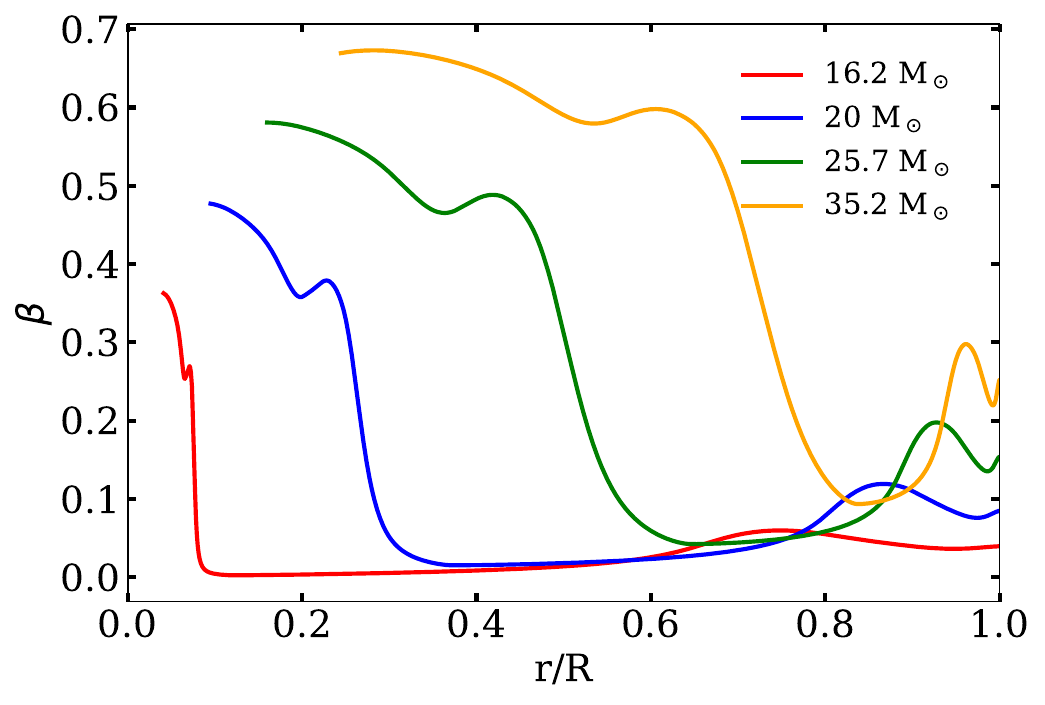}
\end{minipage}
\caption{Density (left) and the ratio of gas pressure to total pressure $\beta$ (right) as a function of relative radius for selected envelope models.}
\label{fig2:rho_beta}
\end{figure}

The modal diagram exhibits a variety of mode crossings and mode couplings which are indicative of the presence of strange modes and associated instabilities. Such modes are not driven by the $\kappa$-mechanism and they are known to occur in stellar models where the luminosity-to-mass (L/M) ratio is greater than $\approx10^4$ in solar units \citep{glatzel1999}, which holds for all the models considered here. Mode interaction and the number of instabilities appears to be more frequent as the L/M ratio increases towards lower masses. This may be due to the change of stellar structure with mass. In Fig.\ref{fig2:rho_beta}, the density as well as the ratio of the gas pressure to the total pressure ($\beta$) is shown for selected models. The density profiles show a distinct core-envelope structure (with an extended low density envelope) which becomes more significant as the mass decreases. The density decreases overall for lower mass models. As a result, the ratio $\beta$ is also affected. For the 16.2 M$_\odot$ model, the value of $\beta$ is low ($<0.1$) for a much larger fraction of the envelope than for the 35.2 M$_\odot$ model. Thus, as the mass of models decrease, they are more likely to become dominated by radiation pressure. These conditions i.e. high L/M ratio, pronounced core-envelope structure and dominant radiation pressure cause strongly non-adiabatic behaviour which appears to favour the excitation of strange mode instabilities \citep{glatzel1994}. 

\subsection{Non-linear simulations} \label{nonlinear}

\begin{figure}
\centering

\begin{minipage}{0.48\textwidth}
    \centering
    \includegraphics[width=\linewidth]{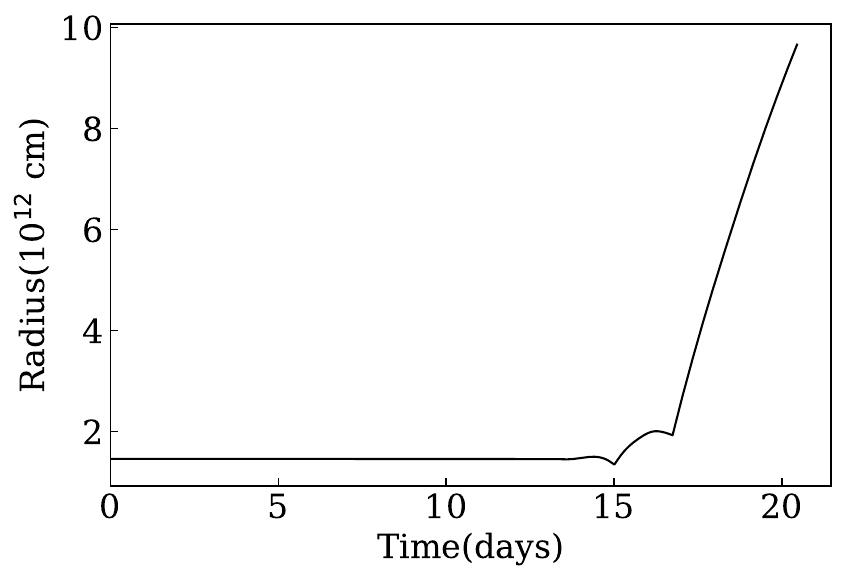}
\end{minipage}
\hfill
\begin{minipage}{0.48\textwidth}
    \centering
    \includegraphics[width=\linewidth]{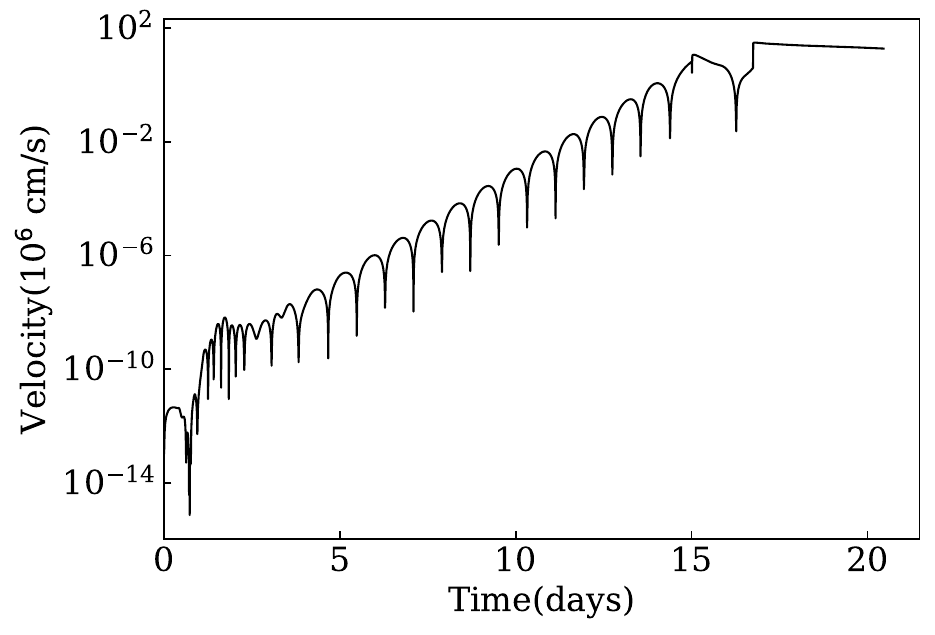}
\end{minipage}
\caption{Evolution of instability into non-linear regime for a 20 M$_\odot$ model: radius (left) and velocity (right) at outermost grid point as a function of time.}
\label{fig3:20m_nonlinear}
\end{figure}

\begin{figure}
\centering

\begin{minipage}{0.32\textwidth}
    \centering
    \includegraphics[width=\linewidth]{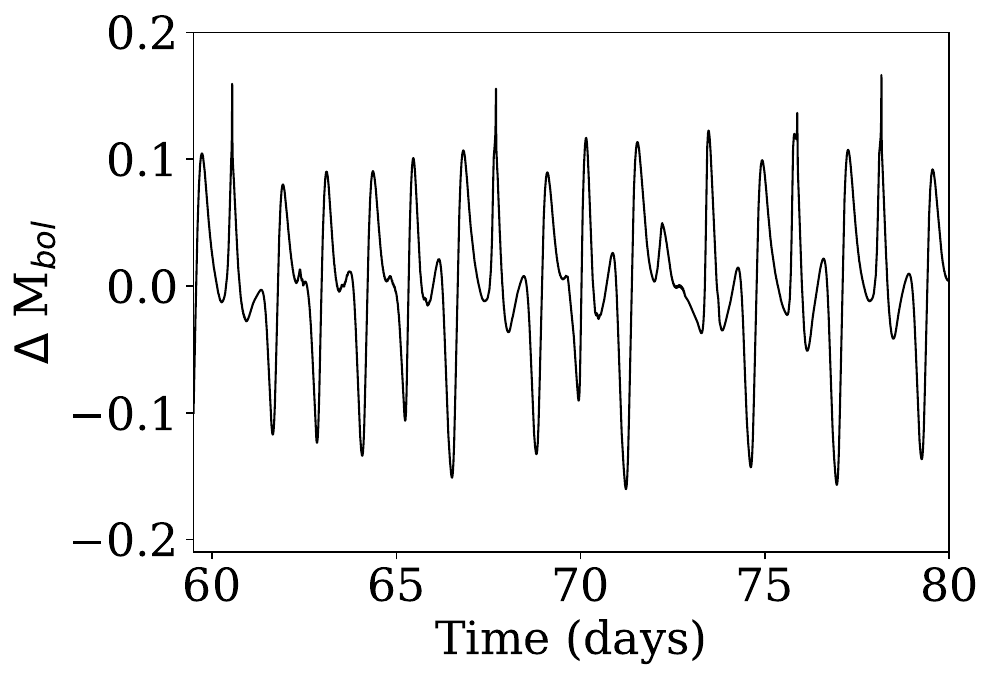}
\end{minipage}
\hfill
\begin{minipage}{0.32\textwidth}
    \centering
    \includegraphics[width=\linewidth]{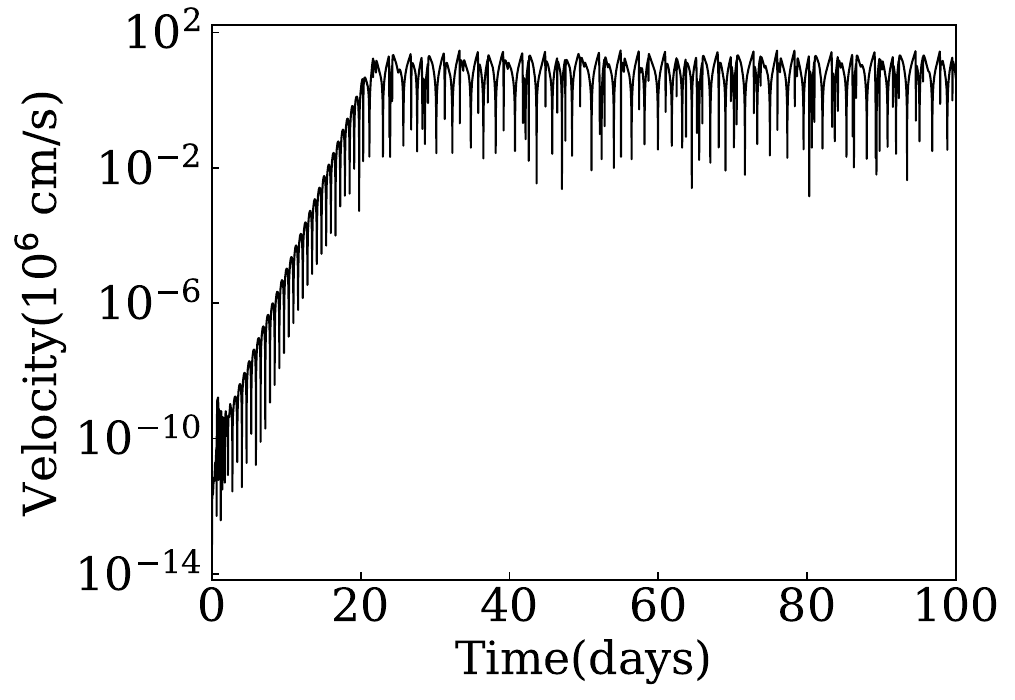}
\end{minipage}
\begin{minipage}{0.32\textwidth}
    \centering
    \includegraphics[width=\linewidth]{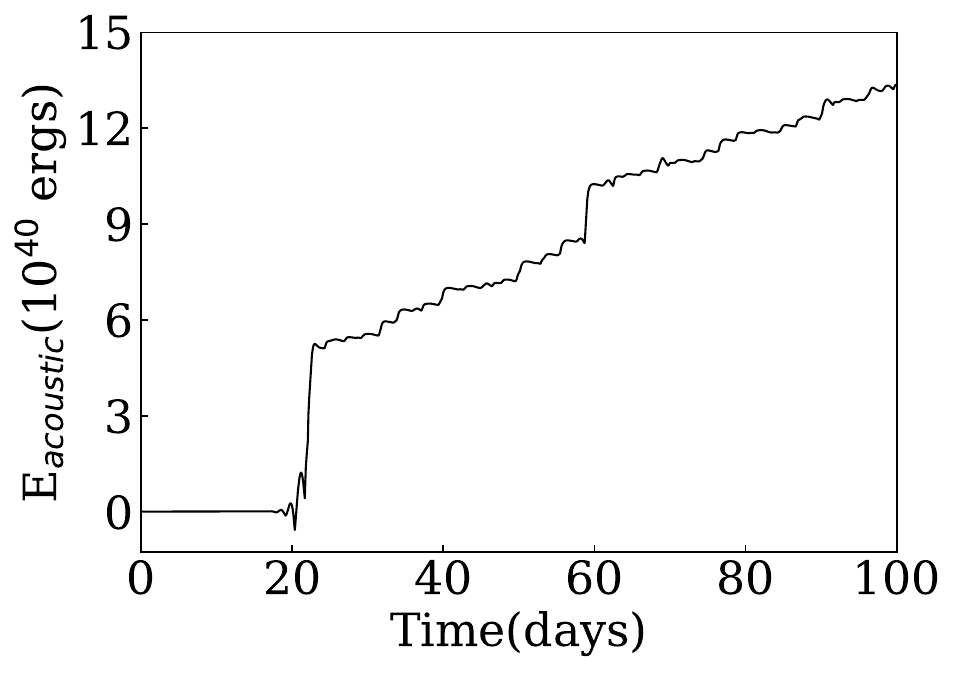}
\end{minipage}
\caption{The non-linear evolution of an instability for 25.7 M$_\odot$ model: variation of bolometric magnitude and velocity at outermost grid point and the acoustic luminosity as a function of time.}
\label{fig4:25.7m_nonlinear}
\end{figure}

The preliminary results of non-linear simulations of two unstable models of 20 M$_\odot$ and 25.7 M$_\odot$ for HD 152249 are presented. Fig.\ref{fig3:20m_nonlinear} shows the variation of the radius and velocity at the outermost grid point for the 20 M$_\odot$ model. The radius is considerably inflated in the non-linear regime. Associated with this is a sharp drop in temperature to below 6000 K and due to unavailability of further opacity data, the simulation was terminated. The velocity reaches a maximum value of $\approx300$ km\,s$^{-1}$ which is 51\% of the escape velocity of the model. For the 25.7 M$_\odot$ model, the variation of the bolometric magnitude and velocity at the outermost grid point as well as time-integrated acoustic luminosity as a function of time are given in Fig.\ref{fig4:25.7m_nonlinear}. The instability here leads to finite amplitude pulsation with no well-defined period. Velocity amplitude shows clear linear phase of exponential growth and saturates in the non-linear regime at $\approx200$ km\,s$^{-1}$. 
The increase of the non-monotonic acoustic luminosity with time on average indicates that the outgoing flux exceeds the incoming flux over one cycle of pulsation, thus transferring acoustic energy to the atmosphere. Hence, a pulsationally driven mass loss seems to be indicated \citep[see e.g.,][]{grott2005,yadav2016}.

\section{Discussion and Conclusion} \label{conclusion}

Linear radial stability analysis has been carried out for models of the O-type supergiant HD 152249. Several modes with growth rates comparable to the dynamical timescale are found to be unstable in the considered models; there seems to be evidence for candidate strange mode instabilities in models of lower masses. However, a Non-Adiabatic-Reversible (NAR) approximation \citep{gautschy1990b} is required to confirm whether they are in fact strange modes, which is beyond the scope of the current study. Periods ranging from 0.3 - 5 d seem to be associated with the unstable modes. Preliminary non-linear simulations carried out for two models with strong growth rates ($\sigma_i<-0.1$) indicate that the instabilities lead to envelope inflation and finite amplitude irregular pulsation and may induce mass loss. 
The non-linear velocity amplitudes however seem to be larger compared to those observed in massive O-type stars \citep{fullerton1996}. 
The present study, restricted to radial pulsations, is a preliminary attempt to explore strange mode instabilities and their outcomes in models of O-type supergiants.
Both theory and observations however suggest that such models should also exhibit non-radial instabilities \citep{dejong1999, mehren1996}. A theoretical study in this direction is intended. Long term photometric and spectroscopic observations using Belgo-Indian Network for Astronomy and Astrophysics (BINA) facilities will be useful to understand the variability in this and other O-type stars and compare with theoretical predictions.



\begin{acknowledgments}
SP acknowledges financial support from the DST-INSPIRE Fellowship (IF220167).
\end{acknowledgments}

\begin{furtherinformation}

\begin{orcids}

  \orcid{0000-0001-8262-2513}{Abhay Pratap}{Yadav}
  \orcid{0009-0000-3108-8744}{Sugyan}{Parida}
  \orcid{0009-0007-1545-854X}{Santosh}{Joshi}
 
\end{orcids}

\begin{authorcontributions}
All authors have contributed significantly.
\end{authorcontributions}

\begin{conflictsofinterest}
The authors declare that there is no conflict of interest.
\end{conflictsofinterest}

\end{furtherinformation}



%

\bibliographystyle{bullsrsl-en}

\bibliography{extra.bib}

\end{document}